\documentclass{elsart}

\usepackage{amssymb}
\usepackage{epsfig}
\usepackage[square,comma,sort&compress]{natbib}

\def\pT{p_{{}_\mathrm{T}}}

\date{28~September~2001}

\begin{document}

\begin{frontmatter}

\title{Radiative decay width of the
$\mathbf{a}_\mathbf{2}$(1320)${}^\mathbf{-}$
meson}

The SELEX Collaboration
\author[Protvino]{V.V.~Molchanov\corauthref{cor}},
\corauth[cor]{Corresponding author.}
\ead{molchanov@mx.ihep.su}
\author[PNPI]{G.~Alkhazov},
\author[PNPI]{A.G.~Atamantchouk\thanksref{tra}},
\author[ITEP]{M.Y.~Balatz\thanksref{tra}},
\author[PNPI]{N.F.~Bondar},
\author[Rochester]{D.~Casey},
\author[Fermi]{P.S.~Cooper},
\author[Flint]{L.J.~Dauwe},
\author[ITEP]{G.V.~Davidenko},
\author[MPI]{U.~Dersch\thanksref{trb}},
\author[ITEP]{A.G.~Dolgolenko},
% ...
\address[Fermi]{Fermilab, Batavia, IL 60510, U.S.A.}
\address[Protvino]{Institute for High Energy Physics, Protvino, Russia}
% ...
\address[ITEP]{Institute of Theoretical and Experimental Physics, Moscow, Russia}
\address[MPI]{Max-Planck-Institut f\"ur Kernphysik, 69117 Heidelberg, Germany}
% ...
\address[PNPI]{Petersburg Nuclear Physics Institute, St. Petersburg, Russia}
% ...
\address[Flint]{University of Michigan-Flint, Flint, MI 48502, U.S.A.}
\address[Rochester]{University of Rochester, Rochester, NY 14627, U.S.A.}
% ...
\thanks[tra]{deceased}
\thanks[trb]{Present address: Infinion, M\"unchen, Germany}
\thanks[trc]{Now at Imperial College, London SW7 2BZ, U.K.}
\thanks[trd]{Now at Instituto de F\'{\i}sica da Universidade Estadual de Campinas, UNICAMP, SP, Brazil}
\thanks[tre]{Now at Physik-Department, Technische Universit\"at M\"unchen, 85748 Garching, Germany}
\thanks[trf]{Present address: The Boston Consulting Group, M\"unchen, Germany}
\thanks[trg]{Now at Instituto de F\'{\i}sica Te\'orica da Universidade Estadual Paulista, S\~ao Paulo, Brazil}
\thanks[trh]{Present address: Lucent Technologies, Naperville, IL}
\thanks[tri]{Present address: SPSS Inc., Chicago, IL}
\thanks[trj]{Now at University of Alabama at Birmingham, Birmingham, AL 35294}
\thanks[trk]{Present address: Imadent Ltd.,\ Rehovot 76702, Israel}
\thanks[trl]{Present address: DOE, Germantown, MD}
\thanks[trm]{ Present address: Siemens Medizintechnik, Erlangen, Germany}
\thanks[trn]{Present address: Deutsche Bank AG, Eschborn, Germany}

\begin{abstract}
Coherent $\pi^+\pi^-\pi^-$ production in the interactions
of a beam of $600\,\mathrm{GeV}$
$\pi^-$ mesons with C, Cu and Pb nuclei has been studied
with the SELEX facility (Experiment E781 at Fermilab).
The $a_2(1320)$ meson signal has been detected
in the Coulomb (low~$q^2$) region.
The Primakoff formalism used to extract radiative decay width of this meson
yields
$\Gamma=284\pm25\pm25\,\mathrm{keV}$,
which is the most precise measurement to date.
\end{abstract}

\begin{keyword}
Radiative decay \sep Primakoff effect \sep a2(1320)
\PACS 13.40.Hq \sep 13.60.Le \sep 14.40.Cs
\end{keyword}

\end{frontmatter}

\maketitle

\section{Introduction}

Radiative decays of mesons and baryons,
as well as other electromagnetic processes,
are important tools for studying
internal structure of these particles
and for testing unitary symmetry schemes and quark models of hadrons.
Such processes, which are governed by interactions
of real and virtual photons with electric charges of quark fields,
make it possible to obtain unique information
about the quark content of hadrons
and about certain phenomenological parameters of hadrons
(magnetic and electric transition moments,
form factors, polarizabilities,~etc).
The underlying processes are simpler to analyze than purely hadronic phenomena,
and can play an important role in testing
chiral, bag, string and lattice models of hadrons.

Direct observation and study of rare radiative decays of hadrons
of the type $a \to h+\gamma$
is often very difficult to carry out
because of high background from $a \to h+\pi^0(\eta)$,
$\pi^0(\eta)\to2\gamma$ decays,
with one lost photon,
and other hadronic processes with $\pi^0(\eta)$ production.
An alternative way to investigate such decays
in coherent production 
in the Coulomb field of atomic nuclei
was proposed initially by Primakoff, Pomeranchuk and
Shmushkevich~\cite{primakoff-1951-prim,pomeranchuk-1961-prim}:
\begin{equation}
h + (A,Z) \to a + (A,Z)
\label{re-coulomb}
\end{equation}
The cross section for such reactions
(which is usually referred to as Primakoff production)
is proportional to the radiative decay width~$\Gamma(a\to{}h+\gamma)$.
It follows that by measuring the absolute cross section
of the Coulomb contribution to Reaction~(\ref{re-coulomb}),
it is possible to determine the radiative width~$\Gamma(a\to{}h+\gamma)$.
Detailed description of this method
and its comparison with possibilities of direct radiative decay studies
can be found in many review
papers~\cite{review},
which also contain results of previous experiments
at high energies.
It must be stressed that determination
of the radiative width $\Gamma(a\to{}h+\gamma)$
in Reaction~(\ref{re-coulomb})
is theoretically straightforward,
while dependence on nuclear structure is minimal at high energies.
In that sense Primakoff technique can be considered
as a direct measurement of the radiative decay width,
in contrast to such methods,
as fit of photoproduction cross section to one pion exchange model,
employed in the first
$\Gamma(a_2\to{}\pi+\gamma)$ measurement~\cite{may-1977-etapi}.
Certainly, analysis of Reaction~(\ref{re-coulomb})
must take into account contributions due to strong interactions
and various interference effects. Usually these are small
and tend to decrease with energy.


In this Letter, we present measurements
of the width for the radiative decay $a_2(1320)^-\to\pi^-+\gamma$
in the Coulomb production reaction
\begin{equation}
\label{re-coulomb-a2}
\begin{array}{rcl}
\pi^- + (A,Z) & \to & a_2(1320)^- + (A,Z)\\
& & \kern0.36em
    \hbox{\vtop to0pt{\vss\hbox{%
        \vrule height18pt depth-3pt}}}
    \kern-0.36em
    \to\pi^+\pi^-\pi^-
\end{array}
\end{equation}
on C, Cu and Pb nuclei at a beam energy
of approximately~$600\,\mathrm{GeV}$
in an experiment using the SELEX spectrometer (E781) at Fermilab.
Preliminary results of this study
were published previously~\cite{kubarovsky-1999-ichep98}.

\section{Experimental apparatus}

The SELEX facility~\cite{smith-1997}
is a forward magnetic spectrometer
with scintillation counters and hodoscopes, proportional
and drift chambers, silicon microstrip beam and vertex detectors,
additional downstream microstrip stations in the beam region,
three lead glass photon detectors, a hadron calorimeter,
two transition radiation detectors~(TRD),
and a multiparticle RICH counter.

The experiment was designed mainly to study
production and decays of charm baryons in a hyperon
beam~\cite{russ-physics}.
It emphasized the forward
($x_{{}_\mathrm{F}}>0.1$) region and, consequently, had high
acceptance for exclusive low multiplicity processes.
Studies of Coulomb production were performed in parallel
with the main charm-physics program and several other measurements.
This imposed certain limitations
on the trigger, geometry and choice of targets.
We report studies
using a negative hyperon beam consisting of
${\simeq}\,50\%\;\Sigma^-$ and ${\simeq}\,50\%\;\pi^-$.
The average beam momentum for pions was~$610\,\mathrm{GeV}$.
For $a_2(1320)$ Coulomb production,
the basic process corresponds to the coherent reaction:
\begin{equation}
\pi^- + A \to \pi^+\pi^-\pi^- + A
\label{re-3pi}
\end{equation}
This was singled out with the help of a special exclusive trigger.
This trigger used scintillation counters to define beam time and
to suppress interactions upstream of the target.
Pulse height in the
interaction counters was used to select events with exactly
three charged tracks downstream of the target.
The trigger hodoscope, which was located after two analyzing magnets,
also required three charged tracks.
Finally, to reduce the background trigger rate to an acceptable level,
the aperture was limited by veto counters,
which had little effect on efficiency for Reaction~(\ref{re-3pi}).
A segmented target with 2~Cu and 3~C foils, each separated
by $1.5\,\mathrm{cm}$, was used for most of the data taking.
A thin Pb target, which is important for the study
of Coulomb production, was used only during brief periods of running
because of the deleterious impact on charm measurements.

Only part of the SELEX facility
was needed for the study of Reaction~(\ref{re-3pi}).
The beam transition-radiation detector provided reliable separation
of~$\pi^-$ from~$\Sigma^-$.
Silicon strip detectors (most of which had $4\,\mu\mathrm{m}$
transverse position resolution) measured parameters of the beam
and secondary tracks in the target region. After deflection by
analyzing magnets, tracks were measured in 14 planes
of $2\,\mathrm{mm}$ proportional wire chambers.
The absolute momentum scale was calibrated using the $K^0_\mathrm{S}$ decays.
Three-pion mass resolution in the $a_2(1320)$ region was~$14\,\mathrm{MeV}$.
A special on-line filter was used
to reduce the number of exclusive events written to tape.
Originally, this selected events that had
at least one secondary reconstructed track,
but it was modified subsequently to require at least two segments after
the analyzing magnets. Very loose criteria were imposed
on the number of hits in the tracking detectors to control processing time.
All these requirements were not very restrictive, and are expected
to have only minor effect on the process of interest.

\section{Data analysis}

Events for Reaction~(\ref{re-3pi})
were selected by requiring a reconstructed beam track and
three charged tracks in the final state.
These tracks were required to form a good vertex
in the vicinity of one of the targets.
The beam particle had to be identified as a pion by the beam TRD.
However, there was no such requirement
for the produced particles.
To suppress inclusive ($\pi^+\pi^-\pi^-+X$) background,
the energy sum of the observed particles
was required to be within $\pm17.5\,\mathrm{GeV}$ of the beam energy.
For further supression of these events,
the most upstream photon detector was used as a guard system,
requiring that any registered energy be less than $2\,\mathrm{GeV}$.
The number of events selected for Reaction~(\ref{re-3pi})
for different targets, and other information of interest,
is summarized in Table~\ref{tab-results}.

\begin{figure}[!s]
\begin{center}
    \epsfbox{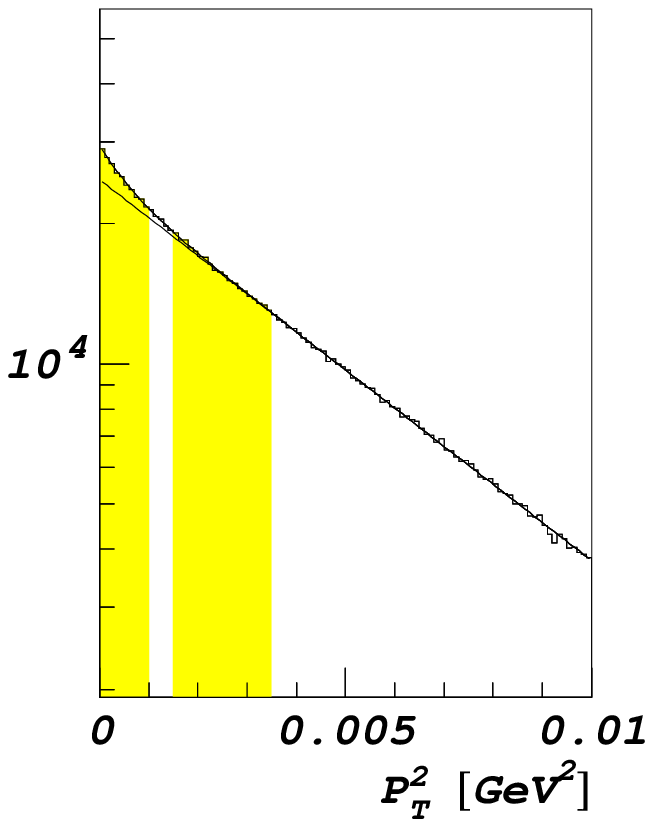}
\end{center}
\caption{Transverse momentum distribution for Reaction~(\ref{re-3pi})
         on a Cu target.}
\label{fig-pt2}
\end{figure}

Most of the ensuing analysis will be described
using the data from the copper target.
The distribution in the square of the transverse momentum~($\pT^2$)
of the $3\pi$-system in Reaction~(\ref{re-3pi})
is shown in~Fig.~\ref{fig-pt2}.
This distribution can be fitted by the sum of two falling exponentials,
one with slope parameter~$b_1\approx180\,\mathrm{GeV}^{-2}$,
which is characteristic of coherent diffractive production on a copper nucleus,
and the other with a slope parameter~$b_2\sim1500\,\mathrm{GeV}^{-2}$,
which is consistent with the estimation for Coulomb production
folded in with the experimental resolution in transverse momentum.
Data for C and Pb targets exhibit similar behavior (not shown),
which establishes the presence of Coulomb production
in Reaction~(\ref{re-3pi}) for all three targets.

\begin{figure}[!s]
\begin{center}
    \epsfbox{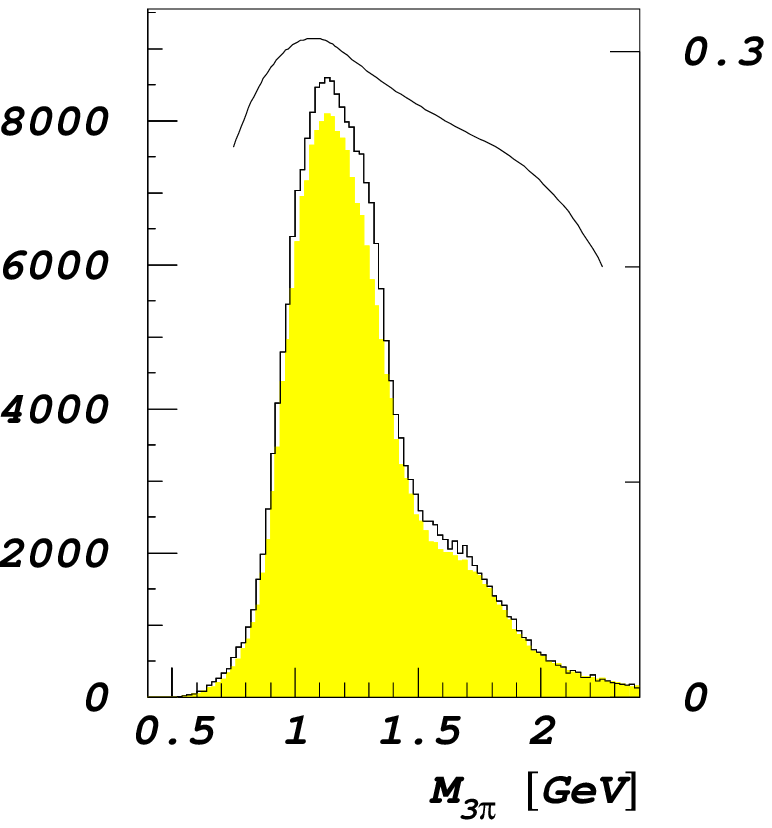}
\end{center}
\caption{Mass distribution for events with $\pT^2<0.001\,\mathrm{GeV}^2$
         (histogram)
         and $0.0015<\pT^2<0.0035\,\mathrm{GeV}^2$, after normalization
         for background subtraction
         (shaded) according to~Fig.~\protect\ref{fig-pt2}.
         The curve shows the efficiency for observing
         a $\rho\pi$ in a $1^+S0^+$ wave,
         which is dominant in the shown mass spectrum.}
\label{fig-mass_bands}
\end{figure}

\begin{figure}[!s]
\begin{center}
    \epsfbox{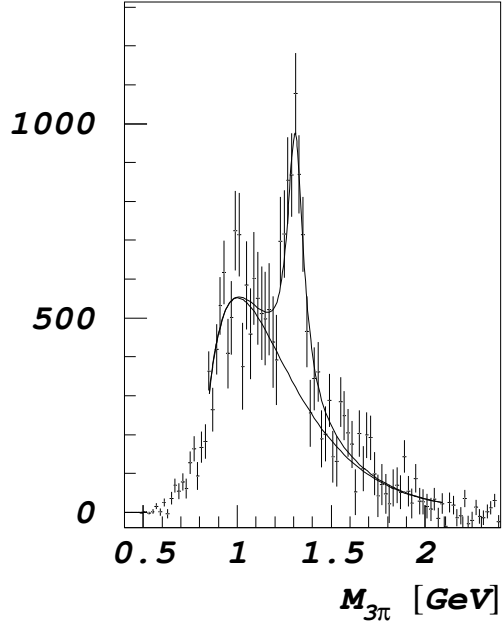}
\end{center}
\caption{$M_{3\pi}$ mass distribution for the Cu target
         after subtraction of diffractive background.
         The curve shows fit with a sum
         of pure Coulomb contribution and smooth background.}
\label{fig-mass_cu}
\end{figure}

\begin{figure}[!s]
\begin{center}
    \epsfbox{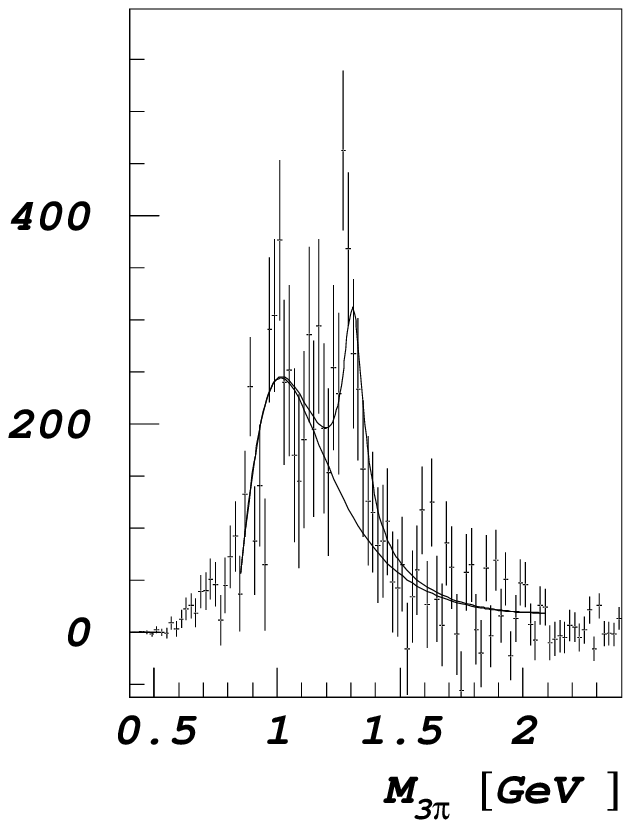}
\end{center}
\caption{The same as Fig.~\protect\ref{fig-mass_cu}, but for C target.}
\label{fig-mass_c}
\end{figure}

\begin{figure}[!s]
\begin{center}
    \epsfbox{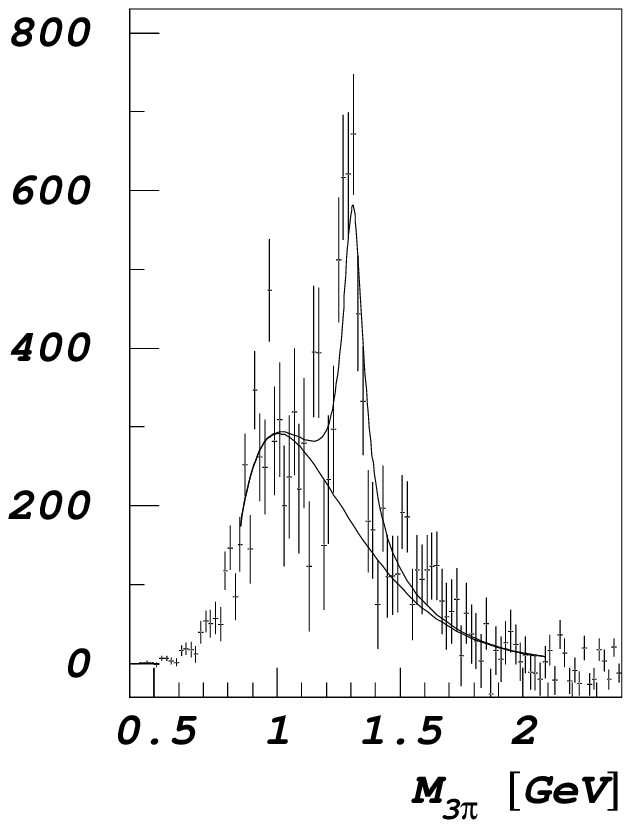}
\end{center}
\caption{The same as Fig.~\protect\ref{fig-mass_cu}, but for Pb target.}
\label{fig-mass_pb}
\end{figure}

Two $\pT^2$ regions are defined for extracting the mass distribution
for the Coulomb production process, as shown in Fig.~\ref{fig-pt2}.
The first one ($\pT^2<0.001\,\mathrm{GeV}^2$)
contains most of the Coulomb contribution,
the second one ($0.0015<\pT^2<0.0035\,\mathrm{GeV}^2$)
has very little of it.
But even the first region is dominated by diffractive production.
The mass spectra $M(3\pi)$ for these two regions
are presented in Fig.~\ref{fig-mass_bands}.
Using results of the fit to Fig.~\ref{fig-pt2},
the mass distribution for events in the second $\pT^2$ region
was normalized to the expected number of diffractive events
in the first region. Then, the mass distribution from the
second region was subtracted from the distribution
for the first $\pT^2$ region.
This type of background subtraction assumes
that the coherent nuclear background at smallest $\pT$
is similar to that at the larger $\pT$ values.
The resulting mass spectrum
is shown in~Fig.~\ref{fig-mass_cu}.
The $a_2(1320)$ signal stands out clearly.
Similar distributions for C and Pb targets are shown
in~Figs.~\ref{fig-mass_c} and~\ref{fig-mass_pb}.
While the observed $a_2$ signal is dominated
by the electromagnetic production mechanism,
there can be contributions to
$a_2$ production from strong interactions (e.g.,~via $f_2$ exchange)
and interference with other mechanisms of $3\pi$ production
(e.g.,~from $a_1(1260)$ Primakoff production).
Corrections for such effects, and the consequent uncertainties,
will be discussed shortly below.


The differential cross section for Coulomb production
of a broad resonance in a pion beam is given by the expression~\cite{%
halprin-1966-prim,faldt-1972-prim-b43,jackson-1964,huston-1986-rho}:
\begin{equation}
\textstyle
\frac{\d \sigma}{\d M \, \d q^2} =
16 \alpha Z^2 (2J+1)
\left( \frac{M}{M^2-m_\pi^2} \right)^3
\frac{m_0^2\Gamma(\pi\gamma)\Gamma(\mathrm{final})}%
{(M^2-m_0^2)^2 + m_0^2\Gamma(\mathrm{all})^2}
\frac{q^2-q^2_\mathrm{min}}{q^4}
|F(q^2)|^2
\label{eq-primakoff_cs}
\end{equation}
where
$\alpha$ is the fine structure constant,
$Z$ is the charge of the nucleus,
$J$ and $m_0$ are spin and mass of the produced resonance,
$M$~is the effective mass of the produced system,
the $\Gamma$ are the decay widths for the corresponding modes,
$q^2$~is the square of the momentum transfer,
and $q^2_\mathrm{min}$~is its minimal value.
At high beam momentum
\begin{equation}
q^2_\mathrm{min}
                 \approx \frac{(M^2-m^2_\pi)^2}{4P^2_\mathrm{beam}}
\end{equation}
At our beam energy, $q^2_\mathrm{min}$ is very small,
and is $\approx2{\cdot}10^{-6}\,\mathrm{GeV}^2$ at the $a_2$~mass.
Consequently,~$q^2 
                   \approx q^2_\mathrm{min}+\pT^2
                   \approx \pT^2$.

The Coulomb form factor $F(q^2)$ in Eq.~(\ref{eq-primakoff_cs})
accounts for the nuclear charge distribution,
initial and final state absorption,
as well as the Coulomb phase.
It was calculated in the framework of the optical model
described in Ref.~\cite{bemporad-1973-prim}.
This model requires knowledge of the
total pion-nucleon cross section~$\sigma$,
and the ratio of real to imaginary parts
of the forward scattering amplitude~$\rho'$,
at the appropriate beam energy.
We used the cross section $\sigma=26.6\,\mathrm{mb}$
determined in the SELEX experiment,
and the extrapolated value of $\rho'=0.12$~\cite{dersch-2000-selex}.
The impact of $F(q^2)$ at these energies is minimal.

The $a_2$ final state was taken to be~$\rho\pi$,
and the total $a_2$ width was parametrized as:
\begin{equation}
\Gamma = \Gamma_0 \frac{m_0}{M} \frac{k}{k_0}
\frac{B_L(kR)}{B_L(k_0R)},
\label{eq-dynamic-width}
\end{equation}
where the $k$ and $k_0$ are center of mass momenta
of $a_2$ decays, off and on resonance, into the corresponding final states.
The $B_L$~are Blatt-Weisskopf centrifugal barrier factors,
as given by von Hippel and Quigg~\cite{hippel-1972}.
The range of interactions $R$ was taken as~$1\,\mathrm{fm}$;
$L$~is the orbital momentum and is equal to~2
for both $\pi\gamma$ and $\rho\pi$ decay modes.

To extract the radiative width $\Gamma(a_2\to\pi\gamma)$
from the Coulomb production of the $a_2(1320)$ meson
given by Eq.~(\ref{eq-primakoff_cs}),
requires an absolute normalization of the cross section.
This means taking into account luminosity of the exposure
and efficiency, which includes trigger, acceptance, reconstruction,
as well as effects of transverse momentum resolution.
The most difficult and uncertain procedure arises from the
evaluation of the trigger performance.
This is because of accidental veto rates,
uncertainties in the discrimination of analog amplitudes,
and other factors that varied during the run.
That is why we chose to normalize the measurement
to the three-pion diffractive production process,
which dominates Reaction~(\ref{re-3pi})
in the region of $q^2\lesssim0.4A^{-2/3}\,\mathrm{GeV}^2$.
As far as the trigger is concerned,
both Coulomb and diffractive production have the same kinematics,
thus, in such an analysis, all trigger and luminosity uncertainties cancel.

Our preliminary result~\cite{kubarovsky-1999-ichep98}
relied on a normalization to the diffractive cross sections
measured by the E272 experiment~\cite{zielinski-1983}.
But these data were obtained under different experimental conditions
($\pi^+$ beam with an energy of $200\,\mathrm{GeV}$)
and had only limited~($\sim15\%$) precision.
Also, we felt it important to avoid any correlation between
our result for the $a_2(1320)$ radiative decay width
and that of the previous E272 measurement~\cite{cihangir-1982-prim}.
Thus, we chose to obtain an independent value
for the diffractive three-pion cross section
in the SELEX experiment,
and normalized our result to the number of events
under the first diffractive exponential of the $\pT^2$ distribution,
as described below.

SELEX had significant periods of running when all the three targets
were employed simultaneously and the trigger did not
distinguish between these targets.
Thus differences in detection efficiency
of Primakoff $a_2$ and diffractive $3\pi$ productions
on different targets
could be described reliably by MC simulations.
Consequently, to obtain a normalization
it was sufficient to measure
the diffractive three-pion cross section
on any of the three target nuclei.

To obtain an absolute normalization,
we used special runs with a so-called ``beam'' trigger.
This trigger employed scintillation counters to define
beam particles and to reject halo, and used no information
from detectors downstream of the targets.
Thus, it selected a completely unbiased set of interactions.
The incident flux was simply
the number of reconstructed beam tracks.
The three-pion mass was confined to $0.8<M(3\pi)<1.5\,\mathrm{GeV}$,
which contains most of the statistics, and for which the acceptance
calculation (to be described later) is very reliable.
Two exposures were analyzed with the beam trigger.
In each, the largest samples
(slightly more than a 1000 diffractive events)
were collected with the carbon target,
which became the natural choice for normalization.
A carbon nucleus is also preferable because it is small,
and therefore diffractive events do not display
an irregular dependence on $\pT^2$ (e.g.,~there is no
large second diffractive maximum),
which could produce additional systematic uncertainties.

The first carbon data sample included short calibration runs
taken at least once a day
under standard experimental conditions.
These indicated that track reconstruction efficiency depended
on beam intensity.
An extrapolation to zero rate provided the result:
$\sigma^{(1)}_\mathrm{diff}=2.39\pm0.14\,\mathrm{mb}$
for the cross section defined above.
The second data set had special stand-alone runs
used to measure
total cross sections with SELEX~\cite{dersch-2000-selex}.
These runs were characterized
by low beam intensity~($\lesssim10\,\mathrm{kHz}$),
use of special targets, and absence of field in the first spectrometer magnet.
The latter led to somewhat higher acceptance,
but worse reconstruction efficiency and momentum resolution.
The measured value of the diffractive cross section
in this data set was
$\sigma^{(2)}_\mathrm{diff}=2.67\pm0.10\,\mathrm{mb}$.

Since the experimental conditions in these independent data sets
were different, it is reasonable to expect that
systematic uncertainties were uncorrelated.
The two measurements were therefore averaged.
Because the $\chi^2$ for the two was~$2.6$
rather than unity,
we followed the usual PDG procedure of scaling the error
by a factor of~$\sqrt{\chi^2}$.
Consequently, the final value used for the normalization on carbon is
${<}\sigma_\mathrm{diff}{>}=2.57\pm0.13\,\mathrm{mb}$.
This result was extrapolated via MC
to Primakoff production on all the targets.

Acceptance and reconstruction efficiencies for all processes were calculated
using a GEANT-based Monte Carlo program~\cite{davidenko-1995-ge781}.
As expected, the efficiency was independent of the $q^2$
for the range relevant to this analysis~($q^2\lesssim0.1\,\mathrm{GeV}^2$).
For Primakoff $a_2$ production, the efficiency was calculated
as a function of mass, with decay kinematics simulated
according to a $\rho\pi$ in a $2^+D1^+$ partial wave
(where $J^PLM^\eta$ corresponds to standard notation~\cite{hansen-1974-pwa},
with $J^P$ being spin and parity of the produced system,
$L$ the relative orbital momentum between the $\rho$ and~$\pi$,
and $M^\eta$ the spin projection and naturality).
For diffractive three-pion production kinematics,
we used $\rho\pi$ in $1^+S0^+$ wave, which is expected
to be dominant~\cite{zielinski-1984-pwa,amelin-1995-pwa}.
The mass was restricted to $0.8<M_{3\pi}<1.5\,\mathrm{GeV}$,
because there is evidence of additional structure
(presumably~$\pi_2(1670)$) at higher mass values.
The shape of the $\rho$-meson was parametrized
using Eq.~(\ref{eq-dynamic-width}).
Comparison of observed and simulated angular and mass distributions
showed good agreement, and thus supported
the assumption about the dominance of the described production mechanism.

To determine the transverse momentum resolution we studied decays of $\Xi^-$,
present in the beam.
We had about 6800 $\Xi^- \to \Lambda\pi^-$, $\Lambda \to p\pi^-$
decays,
with both vertices lying within the target region.
These events are topologically similar to those of Reaction~(\ref{re-3pi}),
and correspond to no momentum transfer~($\pT=0$).
Consequently, the measured momentum transfer
gives the resolution. Comparison of measured values with MC
showed that the transverse momentum resolution is different
for the two transverse $X$ and $Y$ projections,
both in data and MC,
and that the resolution in the MC is better than in the data.
This can be attributed to the idealization of geometry in the MC,
and insufficient detail used in the simulation of detector response and noise.
The difference in quadrature in the resolution between data and MC
$\sqrt{\sigma^2_\mathrm{data}-\sigma^2_\mathrm{MC}}$
was found to be~$\approx5\,\mathrm{MeV}$.
This was used to correct the MC resolutions for $a_2$ production,
which, in general, depended on the data set, target and transverse direction.
The final values vary from $16.2$ to $19.3\,\mathrm{MeV}$,
and have relative uncertainty of~$\approx2\%$.

To obtain the expected shape of the $a_2(1320)$ signal,
Eq.~(\ref{eq-primakoff_cs})
for Coulomb production
was multiplied by efficiency,
convoluted with the $\pT$-resolution,
and integrated over the relevant region of~$\pT^2$.
To check the stability of the result,
we varied the regions of~$\pT^2$ (14 combinations were used)
and employed two different fitting procedures.
In the first procedure,
the subtracted mass distribution,
such as the one shown in~Fig.~\ref{fig-mass_cu}, was fitted with the sum
of a resonance and a smooth background.
In the second procedure,
we fitted the mass distribution
in the first region of~$\pT^2$
(open histogram in~Fig.~\ref{fig-mass_bands}).
To describe background,
we used the distribution from the second (higher) $\pT^2$ region
(shaded histogram in~Fig.~\ref{fig-mass_bands}),
multiplied by a linear function of mass ($a+bM$)
to allow for small changes of shape in the mass spectrum.
Results for different $\pT^2$ regions
and both fitting procedures were similar.
They were used to calculate the average
and to estimate statistical and systematical uncertainties.

The extracted radiative width does not depend strongly
on the form assumed for the shape of the $a_2$ resonance.
This is because the same parametrization must be used both
in fitting the experimental data
and in the expression for the Coulomb production cross section.
In contrast, the total number of $a_2$ events
depends more strongly on the parametrization
because of the relatively large resonance width.
While this number is not used in the analysis
(radiative width is determined directly from the fit),
it provides a measure of the statistical accuracy.
To reduce the dependence on parametrization,
it is customary to count events in a limited mass region.
Such numbers for each target are shown in Table~\ref{tab-results}.

When determining the mass and full width of the $a_2$ from the fit,
we find that they are close to the world average,
while corresponding uncertainties
($\sigma(M)\approx6\,\mathrm{MeV}$ and $\sigma(\Gamma)\approx20\,\mathrm{MeV}$)
are much larger than the world average values~\cite{pdg-2000}.
We consequently fix the mass and width in the fit to their known PDG values.
This has only a small impact on the extracted radiative width.

The $a_2$ signal can be affected by interference
with other $3\pi$ Coulomb production processes.
When intergrated over the phase space,
such interference effects are expected to be small
due to large acceptance of the SELEX apparatus.
One particular case of interest
is Primakoff production of the $a_1(1260)$ meson,
where the dominant decay mode is also~$\rho\pi$
(it is the only meson close in mass to $a_2$
and capable of decaying to $\rho\pi$ and~$\pi^-\gamma$).
Properties of this meson are not well known.
The only measurement of its radiative width to~$\pi\gamma$
is $640\pm246\,\mathrm{keV}$~\cite{zielinski-1984-a1}.
PDG estimation of the full width is $250$--$600\,\mathrm{MeV}$.
Using central values for both widths,
root mean square value of interference effect on the measured
$a_2$ radiative width was estimated to be~$\approx5\%$.
However, data on charge-exchange photoproduction~\cite{condo-1993},
where no evidence of the $a_1$ was found,
while a clear $a_2$ signal was observed,
suggest either an extremely large $a_1(1260)$ total width
or small radiative width to~$\pi\gamma$.
Both possibilities decrease the magnitude of any interference effects.
Given the small value of the described effect,
and significant uncertainties in the properties of the $a_1$ meson,
we do not include this in the systematic
uncertainty on the extracted width.

Because our fitting procedure ignores
strong production of the $a_2(1320)$ meson,
the results of the fit must be corrected for this effect.
It is impossible to correct for interference of the two amplitudes
because the phase difference is not known.
This contributes to a systematic uncertainty
of $\approx4.5\%$ in the analysis.
To describe strong production,
we used the model
developed in Ref.~\cite{bemporad-1973-prim}.
It uses a normalization factor for the $a_2$ production
on a single nucleon $C_\mathrm{S}$,
which must be extrapolated to our energy of $600\,\mathrm{GeV}$.
Production of the $a_2$ meson has been measured
on protons up to an energy of~$94\,\mathrm{GeV}$
(see Ref.~\cite{daum-1980-a2,cleland-1982-a2} and references therein)
and on nuclei at an energy of~$23\,\mathrm{GeV}$~\cite{roberts-1978-pwa}.
We used value $C_\mathrm{S}=1.0\pm0.5\,\mathrm{mb/GeV}^4$,
a large error being assigned to account for the uncertainty
in extrapolation.
Corrections were applied for each combination of $\pT^2$ regions,
and their net effect on the measured radiative width
was estimated as~$\approx3\%$.

The corrected results of the fit for each target,
with their statistical uncertainties,
are shown in Table~\ref{tab-results}.
\begin{table}[!h]
\begin{center}
\begin{tabular}{lrrrl}
Parameter &
\multicolumn{1}{c}{~C} &
\multicolumn{1}{c}{~Cu} &
\multicolumn{1}{c}{~Pb} \\
\hline
Total number of $3\pi$ events & $2.55{\cdot}10^6$ & $1.82{\cdot}10^6$ & $0.55{\cdot}10^6$ \\
Approximate number of $a_2$ events${}^*$ & 1100 & 3700 & 2300 \\
Radiative width $[\mathrm{keV}]$ & 350 & 270 & 291 \\
Statistical uncertainty $[\mathrm{keV}]$ & 121 & 38 & 36 \\
\hline
\multicolumn{5}{p{0.9\hsize}}{${}^*$~This is defined as the
number of resonance events in $1.2$--$1.4\,\mathrm{GeV}$ mass region
in the fits shown in Figs.~\ref{fig-mass_cu}--\ref{fig-mass_pb}.
This differs from the preliminary results
in Ref.~\cite{kubarovsky-1999-ichep98}.}
\end{tabular}
\end{center}
\caption{Characteristics of data on Coulomb $a_2(1320)$
         production on different targets.}
\label{tab-results}
\end{table}
Since most of the factors that contribute to systematic uncertainty
are at least partially correlated for different targets,
the results were averaged over three targets
using only the statistical errors.
Systematic uncertainties include
absolute normalization~($5\%$),
correction for strong $a_2$ production~($1.5\%$),
interference with strong $a_2$ production~($4.5\%$),
transverse momentum resolution~($1.8\%$),
accuracy in $F(q^2)$ calculation~($1\%$),
and uncertainties in the PDG
parameters of the $a_2(1320)$ resonance
mass~($0.35\%$),
width~(3.4\%), and
branching to~$\rho\pi$\footnote{%
        In fact, the relevant branching is $a_2\to3\pi$.
        Possible non-$\rho\pi$ contribution to this decay
        would affect angular distributions and the resonance shape,
        but the effect of this on the measured radiative width
        is negligible.
}~($3.8\%$).
%
%
All sources were added in quadrarture,
and the final combined result is:
\begin{equation}
\Gamma\left[a_2(1320)^-\to\pi^-\gamma\right] =
284 \pm 25 \pm 25 \, \mathrm{keV}
\end{equation}
This is the best measurement to date
(total relative uncertainty of~$12.5\%$).
Comparison with the previous direct measurement~\cite{cihangir-1982-prim}
in the $a_2^+\to\eta\pi^+$ and $K^0_\mathrm{S}K^+$ decay modes,
and with theoretical predictions, is given in Table~\ref{tab-compare}.
\begin{table}[!h]
\begin{center}
\begin{tabular}{l|l}
& $\Gamma\left[a_2(1320)\to\pi\gamma\right]$, $\mathrm{keV}$ \\
\hline
{\em Direct experimental measurements} & \\
SELEX colllaboration (this experiment) & $284\pm25\pm25$ \\
E272 collaboration~\cite{cihangir-1982-prim} & $295\pm60$ \\
\hline
{\em Theoretical predictions} & \\
VDM model~\cite{babcock-1976} & 348 \\
relativistic quark model~\cite{aznauryan-1988} & 324 \\
covariant oscillator quark model~\cite{ishida-1989} & 235 \\
\end{tabular}
\end{center}
\caption{Experimental measurements of
         $\Gamma\left[a_2(1320)\to\pi\gamma\right]$,
         and comparison with theoretical predictions.}
\label{tab-compare}
\end{table}

\subsubsection*{Acknowledgements}

%
%
The authors are indebted to the staff
of the Fermi National Accelerator Laboratory,
and for invaluable technical support from the staffs of collaborating
institutions.
This project was supported in part by Bundesministerium f\"ur Bildung, 
Wissenschaft, Forschung und Technologie, Consejo Nacional de 
Ciencia y Tecnolog\'{\i}a {\nobreak (CONACyT)},
Conselho Nacional de Desenvolvimento Cient\'{\i}fico e Tecnol\'ogico,
Fondo de Apoyo a la Investigaci\'on (UASLP),
Funda\c{c}\~ao de Amparo \`a Pesquisa do Estado de S\~ao Paulo (FAPESP),
the Israel Science Foundation founded by the Israel Academy of Sciences and 
Humanities, Istituto Nazionale di Fisica Nucleare (INFN),
the International Science Foundation (ISF),
the National Science Foundation (Phy \#9602178),
NATO (grant CR6.941058-1360/94),
the Russian Academy of Science,
the Russian Ministry of Science and Technology,
the Turkish Scientific and Technological Research Board (T\"{U}B\.ITAK),
the U.S. Department of Energy (DOE grant DE-FG02-91ER40664 and DOE contract
number DE-AC02-76CHO3000), and
the U.S.-Israel Binational Science Foundation (BSF).

The authors are also grateful to Prof.~B.~Povh
for support of the work connected with
diffractive three-pion production.

\end{document}